\documentclass[twocolumn,showpacs,preprintnumbers,amsmath,amssymb]{revtex4}

\usepackage{graphicx}
\usepackage{dcolumn}
\usepackage{bm}

\begin{document}


\title{Phase-Charge Duality of a Josephson junction in a fluctuating electromagnetic environment}

\author{S. Corlevi$^1$, W. Guichard$^{1,2}$, F.W.J. Hekking$^{2}$, and D.B. Haviland$^1$}
\affiliation{$^1$Nanostructure Physics, Royal Institute of
Technology, 10691 Stockholm, Sweden\\
$^2$University Joseph Fourier and CNRS, B.P. 166, 25 Avenue des
Martyrs, 38042 Grenoble-cedex 09, France\\}


\begin{abstract}
We have measured the current-voltage characteristics of a single
Josephson junction placed in a high impedance environment. The
transfer of Cooper pairs through the junction is governed by
overdamped quasicharge dynamics, leading to Coulomb blockade and
Bloch oscillations. Exact duality exists to the standard
overdamped phase dynamics of a Josephson junction, resulting in a
dual shape of the current-voltage characteristic, with current and
voltage changing roles. We demonstrate this duality with
experiments which allow for a quantitative comparison with a
theory that includes the effect of fluctuations due to finite
temperature of the electromagnetic environment.

\end{abstract}

\pacs{73.23.Hk, 73.40.Gk, 74.50.+r}
\maketitle

Duality often plays a central role in our understanding of the
physical world.  Some physical models, for example the harmonic
oscillator, contain the remarkable property of self duality, where
we can map the model back onto itself in such a way that the role
of physically conjugate degrees of freedom is interchanged.
Another generic model which exhibits self
duality is that of a quantum particle in a cosine potential
coupled to an ohmic heat-bath~\cite{Weiss}. This model also
describes the electrodynamics of a circuit consisting of a single
Josephson junction shunted by a resistor~\cite{footnote}. In this
Letter we present an experimental verification of this self duality.

In the standard theory of the Josephson effect \cite{Josephson,
Likharev}, a current $I$ which is less than the critical current
$I_C$ flows with zero voltage difference across the tunnel
junction, corresponding to a static state where the phase
difference across the junction has no time dependence. This region
of the current-voltage characteristic, known as supercurrent
branch, is well-separated from the dissipative quasiparticle
branch by a voltage of twice the superconducting gap
$V=2\Delta/e$. However, since early work by Ivanchenko and
Zil'berman~\cite{Iv-Zil}, it is known that in the presence of an
impedance $Z_\mathrm{env}(\omega)$ close to a Josephson junction
with small capacitance $C$, phase diffusion will occur, and
therefore dissipative behavior exists also for the supercurrent
with finite, time averaged voltages $\langle V \rangle$  below the
gap. The resulting DC current-voltage characteristic (IVC) of the
junction consists of a supercurrent peak, with a tail at finite
voltages originating from the Josephson oscillations at the
frequency ${f_J=2e\langle V\rangle/h}$. The detailed IVC of the
tunnel junction depends on the charging energy $E_C=e^2/2C$, the
Josephson coupling energy $E_J=\hbar I_{C}/2e$, and the impedance
of the environment $Z_\mathrm{env}(\omega)$. When the impedance is
frequency independent $\Re$e$[Z_\mathrm{env}]=R$, phase diffusion occurs
if $Q=\pi(R/R_{Q})\sqrt{E_{J}/2E_{C}}\ll 1$, where
$R_Q=h/4e^2=6.45~$k$\Omega$ is the quantum resistance. This limit,
usually referred to as overdamped phase dynamics, was analyzed
in~\cite{Iv-Zil}, where the IVC of the junction was calculated in
the presence of thermal fluctuations due to Nyquist noise in the
resistor. Although this analytic result had been known for quite
some time, the supercurrent peak was only recently measured for a
small-capacitance Josephson junction embedded in a carefully
designed low impedance environment, where $R\ll R_Q$ for all
frequencies~\cite{Steinbach}.

The picture developed in~\cite{Iv-Zil} and experimentally verified in~\cite{Steinbach} is based on the classical,
non-linear phase dynamics of the junction and is valid in a low
impedance environment. When $R \simeq R_Q$
quantum fluctuations of the phase become important and
a theoretical description of incoherent tunneling of individual
Cooper pairs can be applied to calculate the junction IVC for an
arbitrary $Z_\mathrm{env}$ when $E_J \ll E_{C}$~\cite{Ingold}. This theory predicts that the
supercurrent peak moves to higher voltages, opening a Coulomb gap
in the IVC of the junction.

Of great interest however is the case when $E_J \geq E_{C}$ and $R\gg R_Q$, usually
referred to as the underdamped case ($Q \gg 1$), where the incoherent tunneling
picture breaks down. Here a situation exactly dual to the above-mentioned classical
overdamped phase dynamics occurs. Based on duality arguments~\cite{Widom} and a
quantitative theory in terms of Bloch bands~\cite{Averin1,Likharev-Zorin}, the IVC of
the junction was predicted to show a voltage peak near zero current corresponding to
Coulomb blockade, and a tail at finite currents corresponding to Bloch oscillations
with frequency $f_{B}=\langle I \rangle /2e$. During each period of the oscillations
exactly one Cooper pair tunnels through the junction. In this picture the junction
circuit is described by the classical, non-linear dynamics of the quasicharge, which
is the conjugate variable to the Josephson phase. In view of duality, when the phase
dynamics is underdamped the corresponding quasicharge dynamics is overdamped. The
quasicharge then evolves in time according to the Langevin equation
$dq/dt=(I_{b}+\delta I) -V/R$. Here, $I_{b}$ is the bias current and $\delta I$ a
random noise component induced by the resistance; the voltage across the junction
$V=dE_{0}(q)/dq$ is given by the derivative of the lowest Bloch energy band (see inset
of Fig.~2). The IVC of a single junction can be analytically calculated from the
Langevin equation for the quasicharge~\cite{Averin1,Likharev-Zorin,Hekking}.

\begin{figure}
\includegraphics[width=\columnwidth]{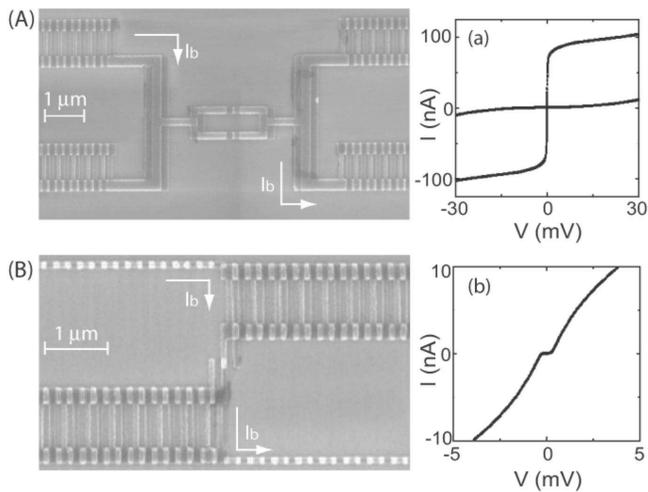}
\caption{ (A) and (B): SEM images of the SQUID and single junction sample configuration. A
symmetric bias $I_{b}$ is applied from the top-left array to the bottom-right array.
(a) and (b): IVC of the current bias lines for the two configurations. (a) from top to
bottom: two SQUID arrays in series in the superconducting ($R_{0}\approx$
50~k$\Omega$) and insulating ($R_{0}\approx$ 50~M$\Omega$) case. (b): two non-SQUID
arrays in series ($R_{0}\approx 10~$M$\Omega$).}
\end{figure}

Few experiments have probed a single Josephson junction in the regime $R\gg R_Q$ due
to difficulties in designing an environment with such high impedance at the frequency
$f_B$ ($\approx 10^{10}$Hz). Small on-chip resistors \cite{Kuzmin, Lotkhov} and
two-dimensional electron gas beneath the junction~\cite{Kycia} have been used to
achieve a high impedance environment. In our experiment, Josephson junction SQUID
arrays are used to bias the junction. The advantage of the SQUID configuration is that
the effective impedance of the environment can be tuned  \emph{in situ} by applying a
magnetic field perpendicular to the SQUID loops.
Although no systematic characterization of the arrays impedance at frequencies of the
order of $f_B$ has been performed, they have been successfully
employed~\cite{Watanabe} to demonstrate how the environment induces Coulomb blockade
and Bloch oscillations in a single Josephson junction in the weak coupling regime
$E_{J}<E_{C}$.

In this paper we present experimental results on single junctions where
$\Re$e$[Z_\mathrm{env}]\gg R_{Q}$.  We have studied a single junction with SQUID geometry (Fig.~1A),
which allows for a systematic study of the IVC of the
same single junction as it is tuned from
strong ($E_{J}>E_{C}$) to weak coupling ($E_{J} < E_{C}$),
with the magnetic field. We also studied a single non-tunable junction (Fig.~1B)
with strong coupling, where the
exact dual of the overdamped Josephson effect is realized.  Here we make
for the first time a detailed quantitative comparison with theory.

The Al/AlOx/Al tunnel junctions are fabricated by double angle evaporation through a
mask patterned by electron beam lithography. The samples are mounted in a RF-tight
copper box and measured in a dilution refrigerator with base temperature of 15~mK. No special cold microwave filters 
were implemented in the cryostat, as the arrays themselves are acting as extremely good filters, protecting the CPT 
from fluctuations generated in the bias circuit. The single junctions are biased
by four Josephson junction arrays and the IVC is measured in a four point
configuration.  One pair of arrays is used to apply a symmetric bias; the current is
measured directly with a current preamplifier (modified Stanford Research System
SR570). The voltage across the single junction is measured through the other pair of
arrays with a high input impedance differential amplifier having an input current of
3~fA (Burr-Brown INA116). Since the impedance of the voltage leads can be in the
G$\Omega$ range, even such small input current can lead to voltages comparable with
the blockade voltage of the single junction, making the measurement of the voltage
over the junction impossible.

In the sample layout presented in Fig.~1A, the single
junction has a SQUID geometry, and it is biased
by four SQUID arrays. Each array consists of 60 SQUIDs with a
nominal area of 0.06~$\mu$m$^{2}$. Each junction of the
central SQUID has an area of $\approx$~0.02~$\mu$m$^{2}$.
The loop of the single junction SQUID is designed to be 10 times larger than the loop area of
the SQUID arrays, enabling a periodic modulation of the
$E_{J}/E_{C}$ of the single SQUID which is incommensurate with the
period of modulation of the environment impedance.
As the magnetic field is increased, the SQUID arrays undergo a
superconducting-insulator quantum phase
transition~\cite{Efetov80}. This results in an increase of the
zero bias resistance $R_{0}$ of the arrays over several orders of
magnitude (from k$\Omega$ up to G$\Omega$). Figure~1a shows the IVC
of the two biasing SQUID arrays at magnetic fields corresponding to the maximum and
minimum value of $R_0$.

Figure 2 shows the IVC of the single junction SQUID at two different values of
magnetic field.  The field values were chosen so that the zero bias
resistance of the biasing leads was
the same for both curves ($R_{0}\approx 10~$M$\Omega$).
However, the $E_{J}/E_{C}$ ratio of the single junction SQUID is
different for the two curves. Curve A in Fig.~2
corresponds to the maximum value of the Josephson
coupling, while in curve B the Josephson coupling
was at a minimum.
We estimate the values of $E_{J}/E_{C}$ for the two curves from the samples parameters.
The charging energy $E_C=e^2/2C=45~\mu$eV  is calculated from the junction area, which
gives a capacitance $C\approx$~1.8~fF assuming a specific capacitance of $45~$fF/$\mu$m$^{2}$.
$E_J$ of the single junction SQUID depends on the magnetic field $B$,
and it can be expressed  as $E_{J}=R_{Q}\Delta/2R_{N} \mid \cos(\pi B /  B_{0})\mid$,
where $R_{N}=2.8~$k$\Omega$ is the normal state resistance of the SQUID, $\Delta \approx 200~\mu$eV
the superconducting gap of Al, and $B_0=16~$G the measured period of the SQUID modulation.
Here we assume that the two junctions in the SQUID are identical.
Thus we calculate $E_{J}/E_{C}=4.5$ at the maximum (curve A).
Experimental uncertainties make
determination much less accurate at the minimum
of the SQUID modulation, where we estimate
$E_{J}/E_{C}\leq 0.2$ (curve B).

\begin{figure}
\includegraphics[width=\columnwidth]{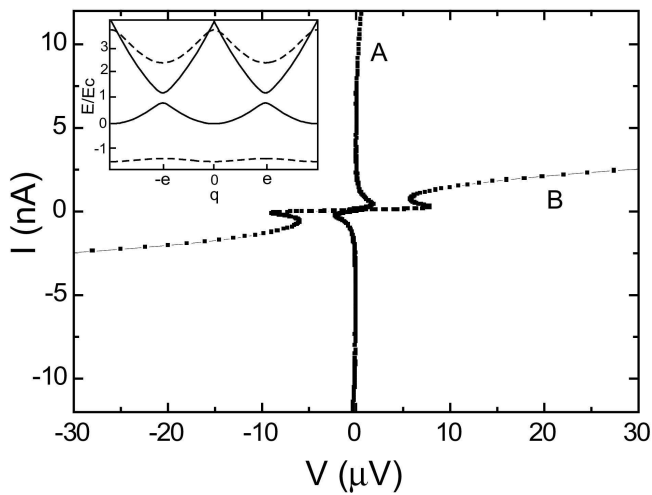}
\caption{IVC of a single junction SQUID at two different values of
magnetic field, corresponding to (A) $E_{J}/E_{C}=4.5$ and (B)
$E_{J}/E_{C}\leq 0.2$. At these values the biasing SQUID arrays
had the same zero bias resistance $R_0 \approx 10~$M$\Omega$. The
inset shows the two lowest energy bands of a single junction
calculated for $E_{J}/E_{C}=2$ (dashed line) and $E_{J}/E_{C}=0.2$
(solid line).}
\end{figure}

Both curves A and B in Fig.~2 show a Coulomb blockade feature,
followed by a back bending region due to the Bloch oscillations.
The differences between these two curves are qualitatively
understood from the Bloch band
theory~\cite{Likharev-Zorin,Schon-Zaikin}. The critical voltage,
or maximum blockade voltage, is determined by the shape of the
lowest energy band, $V_C=\mbox{max}[dE_{0}(q)/dq]$. For
$E_{J}/E_{C}>1$ the lowest energy band becomes very flat (see
inset of Fig.~2), which explains why curve A has a smaller
blockade voltage than curve B.  Furthermore, the gap between the
lowest and first excited energy band strongly depends on
$E_{J}/E_{C}$ (inset Fig.~2). The gap determines the maximum
current, or Zener break-down current $I_Z$, above which Zener
tunneling to higher bands leads to dissipation and increased
voltage across the junction. For $E_J/E_C=4.5$ the experiment
shows $I_Z \simeq 10$~nA, whereas for $E_J/E_C \leq 0.2$, we
measure $I_Z\simeq 0.5$~nA, in qualitative agreement with theory.

\begin{figure}
\includegraphics[width=\columnwidth]{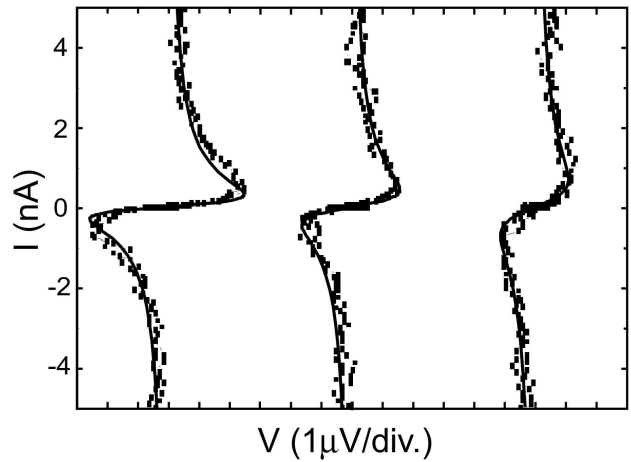}
\caption{\label{fig:epsart}IVC of a single junction in the overdamped quasicharge regime
at different temperature (points) and the theoretical prediction (solid lines). In
the fit V$_{C}=30~\mu$V. From left to right the cryostat (fitting noise) temperatures
are: T=50~mK (160~mK), 250~mK (260~mK), 300~mK (400~mK).}
\end{figure}

This qualitative comparison can be made more quantitative if we
take in to account the effect of fluctuations due to the finite
temperature electromagnetic environment, which cause the measured
blockade voltage to be smaller than the theoretical critical
voltage $V_C$. Using duality arguments, Beloborodov {\it et
al.}~\cite{Hekking} have calculated the IVC of a junction in a
high impedance environment with $E_{J}/E_{C}>1$ in the presence of
thermal fluctuations. In the theory the environment consists of a
resistor ($Z_\mathrm{env}=R$) characterized by gaussian noise with
temperature $T_\mathrm{noise}$. Ref.~\cite{Hekking} assumes the
noise to be white, {\em i.e.}, to have classical correlations,
which is the case as long as $k_B T_\mathrm{noise}
> R_Q eV_C/R$. For our experiment, this condition is satisfied as long as $T_\mathrm{noise} > 5~$mK.

In the experiment, the biasing arrays may generate
more complicated fluctuations, but for the purpose of
comparison we describe the arrays by one parameter, $R$, which represents the most
basic way to characterize their fluctuations.
For a bias current less than $I _{Z}$, the IVC of the junction obtained in~\cite{Hekking}
can be expressed in a form dual to that of~\cite{Iv-Zil,Steinbach}
\begin{equation}
  \langle V\rangle = V_C \Im \mbox{m}\left[\frac{I_{1- i\beta eI_{b}R/\pi}(\beta e V_C/\pi)}{I_{-i \beta
eI_{b}R/\pi}(\beta e V_C/\pi)}
  \right],
   \label{hek}
\end{equation}
where $\beta = 1/k_B T_\mathrm{noise}$ is the inverse of the noise
temperature and $I_{\nu}(z)$ is the modified Bessel function of
argument $z$ and complex order $\nu$.

We have experimentally studied the influence of the cryostat temperature on the
IVC of samples with the layout presented in Fig.~1B.  Here the
single junction under test (area $\approx 0.02~\mu$m$^{2}$) is not
tunable, and it is current biased by two non-SQUID arrays
consisting of 16 junctions with nominal area 0.01$~\mu$m$^{2}$.
The IVC of these arrays (Fig.~1b) shows a strong Coulomb blockade,
providing a good current bias for the single junction. However the
voltage probes are SQUID arrays, allowing the tunability of their
impedance to an optimal value. We have found that the tunability
of the voltage probes is essential for this type of experiment.

\begin{figure}
\includegraphics{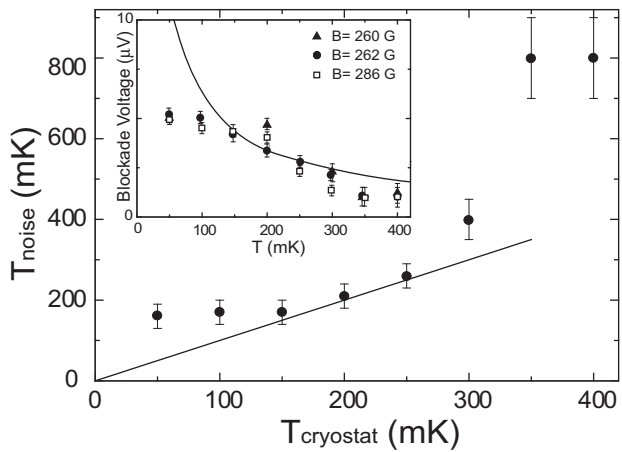}
\caption{Noise temperature as determined from fitting to theory (Fig.~3) plotted vs. the cryostat temperature. Inset: 
Measured blockade voltage for three sets of data (symbols) and calculated blockade voltage for $V_{C}=30~\mu$V (solid 
line) if $T_\mathrm{noise}=T_\mathrm{cryostat}=T$.}
\end{figure}

In Fig.~3 the measured IVCs at fixed magnetic field ($B=262$~G) and various
temperatures are shown together with the theoretical curves
calculated using Eq.~(1). For these samples, the Josephson
coupling of the single junction was large enough to realize the
exact dual of the overdamped Josephson effect. In contrast to
previous works~\cite{Watanabe}, we were able to measure IVCs
showing a complete back bending to a zero voltage current,
allowing for the first time a quantitative comparison between the
theory and experiment. Excellent agreement is achieved using one
value of the parameter $R=150~$k$\Omega$ for all curves, while
adjusting the noise temperature $T_\mathrm{noise}$ for a best fit.
For the theoretical curves the value $V_{C}=30~\mu$V was used,
which compares very well to the estimated critical voltage
$V_{C}=28\pm 7~\mu$V. This value was calculated~\cite{Zorin-priv}
from the energies $E_{J}=270~\mu$eV and $E_{C}\approx 90~\mu$eV,
determined from the junction parameters as previously described
($R_{N}=2~\mbox{k}\Omega$, $C\approx 0.9~$fF). The value
$R=150~$k$\Omega$, which represents the theoretical effective
frequency independent impedance seen by the single junction,
compares rather well with the measured $R_0 \approx 200~$k$\Omega$
of the voltage probes (SQUID arrays) at the particular magnetic
field studied. These arrays exhibit a weaker Coulomb blockade than
the non-SQUID arrays, hence the former determine the impedance
which the junction sees.

Figure 4 shows a comparison between $T_\mathrm{cryostat}$ and
$T_\mathrm{noise}$, where the latter is determined by the fit to theory as shown in Fig.~3. In the
region 175~mK to 250~mK, the two temperatures coincide, but the
noise temperature does not go below 175~mK. This saturation
indicates the existence of residual noise with an effective
temperature around 175~mK in the measurement system. As reported
in~\cite{Steinbach}, excess noise may be a consequence of
inadequate filtering of the measurement leads. Another possible
source of excess noise is shot noise in the current biasing
arrays. These arrays have a strong Coulomb blockade, where charge
transport is discrete and a shot noise is expected. We also find
excess noise when the cryostat is above 250~mK.  This temperature
is close to the estimated odd-even free energy
difference~\cite{Tuominen} for the islands in the arrays, above
which quasiparticle tunneling in the arrays would lead to excess
noise.

Thus we see that noise reduces the measured blockade voltage below
the theoretically predicted maximum critical voltage
$V_{C}=30~\mu$V. To further illustrate this point we plot the
measured blockade voltage vs. $T_\mathrm{cryostat}$ in the inset
of Fig.~4. The solid line is the theoretical prediction if
$T_\mathrm{cryostat}= T_\mathrm{noise}$, independent of the
parameter $R$.  We see that a blockade voltage of the order of 10~$\mu$V
would be expected at a temperature of 50~mK, if the excess noise
could be reduced. It is remarkable that this theory, which uses a
minimal description of the fluctuations in the Josephson junction
arrays, gives the correct order of magnitude for $V_{C}(T$), and
very accurately reproduces the shape of the IVC. A more complex
model, including frequency dependent impedance of the arrays, as
well as additional sources of noise may give better
correspondence, but at the price of many more parameters.

In summary, we have experimentally studied the IVC of a single Josephson junction in a
high impedance environment ($R>R_Q$) in the strong coupling regime ($E_J>E_C$), where
the dual of the classical Josephson effect is realized. We show qualitative agreement
with the Bloch band theory for the same single junction as the ratio $E_J/E_C$ is
tuned with the magnetic field.  By taking in to account the finite temperature of the
electromagnetic environment, we can quantitatively explain the measured data with a
minimal theory, where gaussian fluctuations are described by one parameter.

We acknowledge discussions with H. Grabert, G. Ingold, and M.H. Devoret. This work was
partially supported by the Swedish SSF Center NanoDev, the EU project SQUBIT and
Institut Universitaire de France.



\begin{thebibliography}{19}
\bibitem[{foo()}]{footnote}
\bibinfo{note}{The heat bath of the original model corresponds to a
  reflectionless transmission line. If the line parameters are
  chosen appropriately, it is equivalent to a resistor}.
\end{thebibliography}

\end{document}